\documentclass[aps,pra,twocolumn,reprint,showpacs,amsmath,amssymb,floatfix,superscriptaddress]{revtex4-1}
\usepackage{epsfig}
\usepackage{bm}
\usepackage{color}
\usepackage{array}

\renewcommand{\vec}[1]{\mathbf{#1}}

\newcommand{\ep}{\epsilon}
\newcommand{\vep}{\varepsilon}

\begin{document}

\title{Exciton-phonon relaxation bottleneck and radiative decay of thermal exciton reservoir in two-dimensional materials}

\author{A. O. Slobodeniuk}
\affiliation{Laboratoire National des Champs Magn\'etiques Intenses,
CNRS-UJF-UPS-INSA, 25 rue des Martyrs,
             B.P. 166, 38042 Grenoble, France}

\author{D. M. Basko}
\affiliation{Laboratoire de Physique et Mod\'elisation des Milieux Condens\'es,
Universit\`e de Grenoble-Alpes and CNRS,
25 rue des Martyrs, 38042 Grenoble, France}
\date{\today}

\begin{abstract}
We study exciton radiative decay in a two-dimensional material, taking into account large thermal population in the non-radiative states, from which excitons are scattered into the radiative states by acoustic phonons. We find an analytical solution of the kinetic equation for the non-equilibrium distribution function of excitons in the radiative states. Our estimates for bright excitons in transition metal dichalcogenides indicate a strong depletion of radiative state population due to insufficient exciton-phonon scattering rate at low temperatures.
\end{abstract}

\maketitle

\section{Introduction}

Exciton radiative decay in two-dimensional structures was first
studied for excitons in molecular crystals~%
\cite{Agranovich1966,Aaviksoo1987}.
Later, it attracted much attention in the context of excitons in
semiconductor quantum wells, whose fabrication became possible
due to progress in semiconductor growth techniques~%
\cite{Feldmann1987,Hanamura1988,Andreani1990,Andreani1991,
Damen1990,Deveaud1991,MartinezPastor1993}.
The recent intense studies of monolayer transition metal dichalcogenides
(TMDCs) have lead to a revival of research activity in the radiative
dynamics of two-dimensional excitons, both experimental~%
\cite{Splendiani2010,Mak2010,Korn2011,Shi2013,Lagarde2014,%
Moody2015,Poellmann2015,Koperski2015,Srivastava2015,%
Koirala2016,Dey2016,Robert2016,Selig2016,Jakubczyk2016}
and theoretical~%
\cite{Glazov2014,Gartstein2015,Dery2015,Palummo2015,Wang2016,Selig2016}.

In clean samples, the in-plane momentum $\vec{p}$ is conserved
during the photon emission. Then, only excitons
with small momenta $p\sim\hbar\omega_\mathrm{ex}/c$ can emit
photons (here $\omega_\mathrm{ex}$ is the excitonic resonance
frequency and $c$~is the speed of light). The subsequent dynamics
of the excitonic population and of the emitted light strongly
depends on the exciton distribution over different momentum states.
If the population was created by a resonant optical excitation,
it is initially concentrated in the radiative region, so the
excitons can quickly decay before being scattered into the
non-radiative states. For a non-resonant optical excitation
(excitation energy high above~$\hbar\omega_\mathrm{ex}$), or
electrical pumping (excitons produced by binding free carriers
injected electrically), the excitons may have time to thermalize
before decaying.
Sometimes, both contributions may be seen in photoluminescence~%
\cite{Korn2011,Lagarde2014,Robert2016}.

\begin{figure}
\includegraphics[width=8cm]{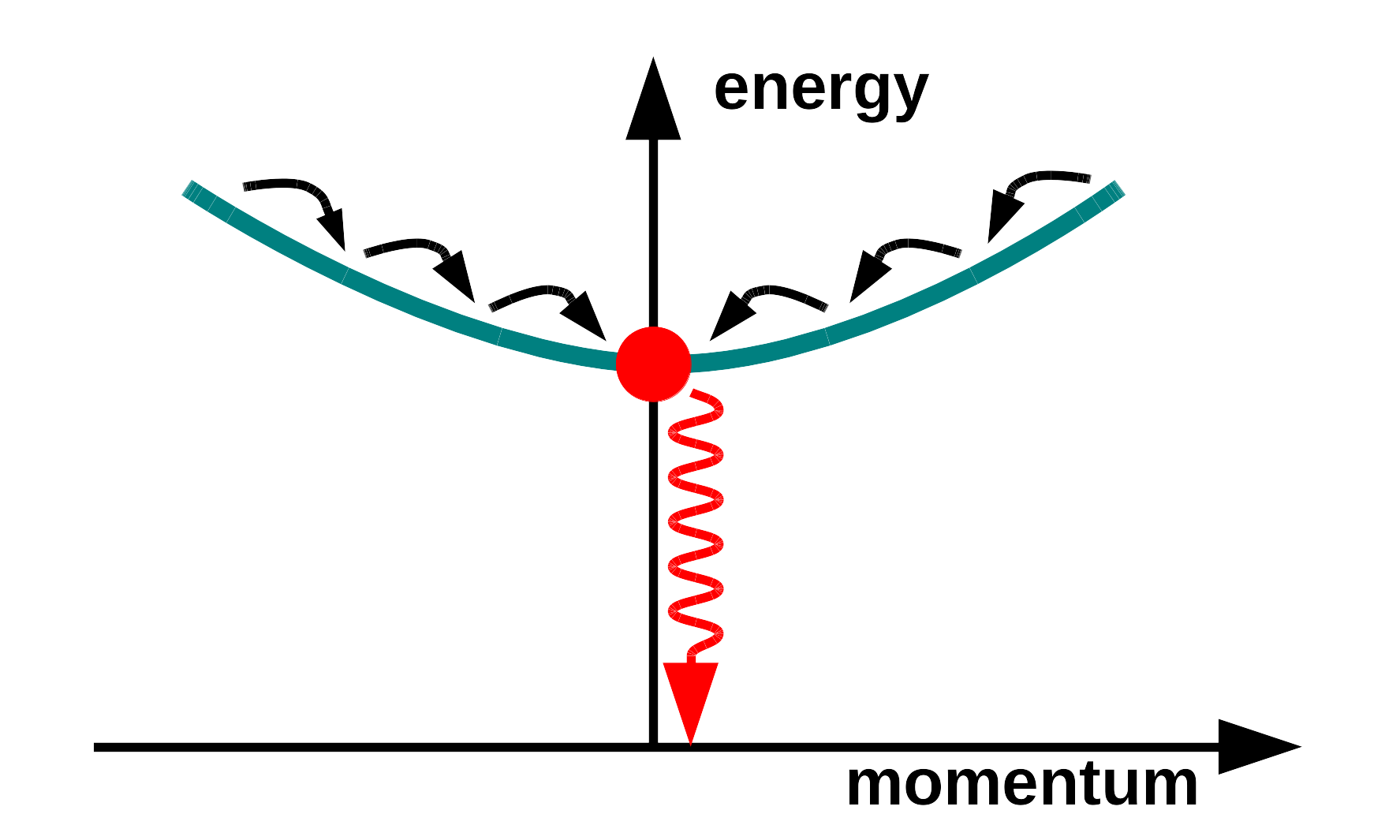}
\caption{\label{fig:relaxation}(Color online)
A schematic representation of the exciton band (solid curve),
with the narrow radiative region shown by the red circle.
Exciton radiative decay by photon emission is represented
by a red wavy arrow, exciton relaxation by phonon emission
is represented by black solid arrows.
}
\end{figure}

If the excitons have thermalized, the excitonic population
extends over a region of momenta, determined by the
temperature~$T$, usually much wider than the radiative region.
Then, to relax the whole population, excitons from the non-radiative
states must be scattered to the radiative region
(Fig.~\ref{fig:relaxation}), so radiative decay of a thermal
excitonic population takes much longer than the recombination
time $1/\Gamma_\vec{p}$ of a radiative state with
small momentum~$\vec{p}$.
If the scattering is fast enough, the exciton distribution remains
thermal, so the population decay rate is given by the simple
thermal average of $\Gamma_\vec{p}$, proportional to
$1/T$~\cite{Andreani1991}.
However, scattering typically slows down at low temperatures,
which results in a depletion of the radiative region, so the overall
population decay rate is slower than the thermal average.
Numerical solution of the Boltzmann equation for excitons
scattered by acoustic phonons in a GaAs quantum well has
shown the importance of this relaxation bottleneck effect~%
\cite{Piermarocchi1996,Thranhardt2000}.

In this paper, we revisit the problem of competition between the
exciton radiative decay and the scattering by acoustic phonons,
assuming the latter to be the dominant source of scattering.
We do not include scattering by optical phonons, assuming the
excitonic temperature to be much lower than the optical phonon
energy.
We also assume the exciton density to be sufficiently low, so
that exciton-exciton scattering and annihilation is inefficient.
We consider a clean undoped sample, so exciton scattering by
impurities and free carriers can also be neglected.
Under these conditions, acoustic phonons can dominate the exciton
scattering.

Below, we show that the Boltzmann equation describing the
exciton radiative decay and scattering by acoustical phonons
has a remarkably simple
analytical solution, whose validity is guaranteed by the
same separation of momentum scales that creates the problem:
the large width of the excitonic thermal distribution as
compared to that of the radiative region.
This makes the exciton distribution in the non-radiative
reservoir insensitive to what happens in the radiative region,
and enables us to find the overall population decay rate
in the bottleneck regime
[Eqs.~(\ref{avGammaNarrow=}) and (\ref{avGammaBroad=}), for
$\Gamma_\vec{p}$ being smaller or larger than the typical
phonon frequency, respectively].
Applying our results to bright excitons in TMDCs, we find a
strong bottleneck effect, leading to nanosecond population
decay times at $T\sim{10}-100$~K (neglecting non-radiative decay).

\section{Single exciton band}

We start by considering
the simplest case of a non-degenerate exciton band
with the mass $m_\mathrm{ex}$ and the parabolic dispersion
$\hbar\omega_\mathrm{ex}+p^2/(2m_\mathrm{ex})$,
weakly coupled to two-dimensional acoustic phonons
and to three-dimensional photons.
At times longer than the dephasing time, the coherent
polarization can be neglected, and the exciton population
can be characterized by the momentum
distribution function~$f_\vec{p}$. Its time evolution is
described by the kinetic equation:
\begin{equation}\label{kinetic=}
\frac{\partial f_\vec{p}}{\partial t}=
-\Gamma_\vec{p}\,f_\vec{p}
+I^\mathrm{in}_\vec{p}-I^\mathrm{out}_\vec{p}.
\end{equation}
Here $\Gamma_\vec{p}$ is the exciton radiative
decay rate, non-zero for momenta
$p<\hbar{k}_\mathrm{rad}\equiv\sqrt{\vep}\,\hbar\omega_\mathrm{ex}/c$,
where $\vep$ is the dielectric constant of the medium surrounding
the excitonic layer,
for simplicity assumed to be the same on both sides of the layer.
We neglect non-radiative exciton decay,
which can be straightforwardly incorporated.
The last two terms in Eq.~(\ref{kinetic=}),
$I^\mathrm{in}_\vec{p}$ and $I^\mathrm{out}_\vec{p}$, represent
the in- and out-scattering parts of the
exciton-phonon collision integral.
Assuming $f_\vec{p}$ to be non-degenerate,
we write them as
\begin{subequations}\begin{eqnarray}
&&I^\mathrm{in}_\vec{p}=\int\frac{d^2\vec{p}'}{(2\pi\hbar)^2}\,
W_{\vec{p}'\to\vec{p}} f_{\vec{p}'},\label{Iin=}\\
&&I^\mathrm{out}_\vec{p}=
\int\frac{d^2\vec{p}'}{(2\pi\hbar)^2}\,
W_{\vec{p}\to\vec{p}'} f_\vec{p}\equiv\frac{f_\vec{p}}{\tau_\vec{p}}.
\label{Iout=}
\end{eqnarray}\end{subequations}
Here $W_{\vec{p}\to\vec{p}'}$ is the rate of exciton
scattering from the state $\vec{p}$ to the state~$\vec{p}'$,
due to phonon absorption or emission. It includes
the energy-conserving $\delta$~function,
$\delta(p^2/(2m_\mathrm{ex})-(p')^2/(2m_\mathrm{ex})
\pm{u}_\mathrm{s}|\vec{p}-\vec{p}'|)$,
where $u_\mathrm{s}$~is the speed of sound. In Eq.~(\ref{Iout=}),
we also defined the out-scattering time~$\tau_\vec{p}$.
We assume the phonons to be always in equilibrium with
temperature~$T$ (here and below measured in energy units),
then the rates in Eqs.~(\ref{Iin=}),~(\ref{Iout=}) satisfy the
detailed balance condition,
$W_{\vec{p}'\to\vec{p}}/W_{\vec{p}\to\vec{p}'}=
e^{(\ep_{\vec{p}'}-\ep_\vec{p})/T}$. Then, the collision
integral is nullified by the Maxwell-Boltzmann
distribution,
\begin{equation}\label{feq=}
f^\mathrm{eq}_\vec{p}=\frac{2\pi\hbar^2n_\mathrm{ex}}{m_\mathrm{ex}T}\,
e^{-p^2/(2m_\mathrm{ex}T)},
\end{equation}
where $n_\mathrm{ex}=\int{f}_\vec{p}\,d^2\vec{p}/(2\pi\hbar)^2$
is the total exciton density.

Generally, there are several ways to use the kinetic
equation~(\ref{kinetic=}) to study relaxation kinetics.
(i)~One can take some initial condition $f_\vec{p}(t=0)$ and study
its subsequent evolution.
(ii)~One can look for stationary solutions of
Eq.~(\ref{kinetic=}), which must be supplemented by an
exciton generation term. The explicit form of this generation term
would be strongly dependent on the specific experimental situation,
and we prefer to assume that excitation has been performed in the
past, and the excitons have had enough time to thermalize.
(iii)~One can look for decaying solutions of the form
$f_\vec{p}(t)=f_\vec{p}e^{-\gamma{t}}$, where $-\gamma$~is an
eigenvalue of the right-hand side of Eq.~(\ref{kinetic=}),
which is a linear integral operator. Generally, this operator has
several eigenvalues, and we are interested in the smallest one
(by the absolute value). It will also dominate the solution of
the initial value problem~(i) at long times, and can be called
the effective decay rate. Thus, in the following we will focus
on problem~(iii), and look for the effective decay rate~$\gamma$.

The central question is how much the distribution 
in the radiative region is different from the equilibrium one.
This distribution is determined by 
three processes: (i)~exciton radiative decay,
(ii)~exciton scattering from the radiative region to the
non-radiative states with $p>\hbar{k}_\mathrm{rad}$ by phonon
absorption, and (iii)~exciton scattering from the non-radiative
states to the radiative region by phonon emission.
The momenta $\vec{p}'$ of the relevant states in the
non-radiative region are fixed by the energy conservation:
\begin{equation}\label{conservation=}
\frac{(p')^2}{2m_\mathrm{ex}}=\frac{p^2}{2m_\mathrm{ex}}\pm
u_\mathrm{s}|\vec{p}-\vec{p}'|.
\end{equation}
For $\vec{p}$ in
the radiative region, we have $p\ll{p}'$, only the ``+''
sign is allowed, so $\vec{p}'$ must lie in a narrow circular
strip
$|p'-2m_\mathrm{ex}u_\mathrm{s}|<\hbar{k}_\mathrm{rad}$.
It is only from this strip that excitons can scatter
into the radiative region.

\begin{figure}
\includegraphics[width=6cm]{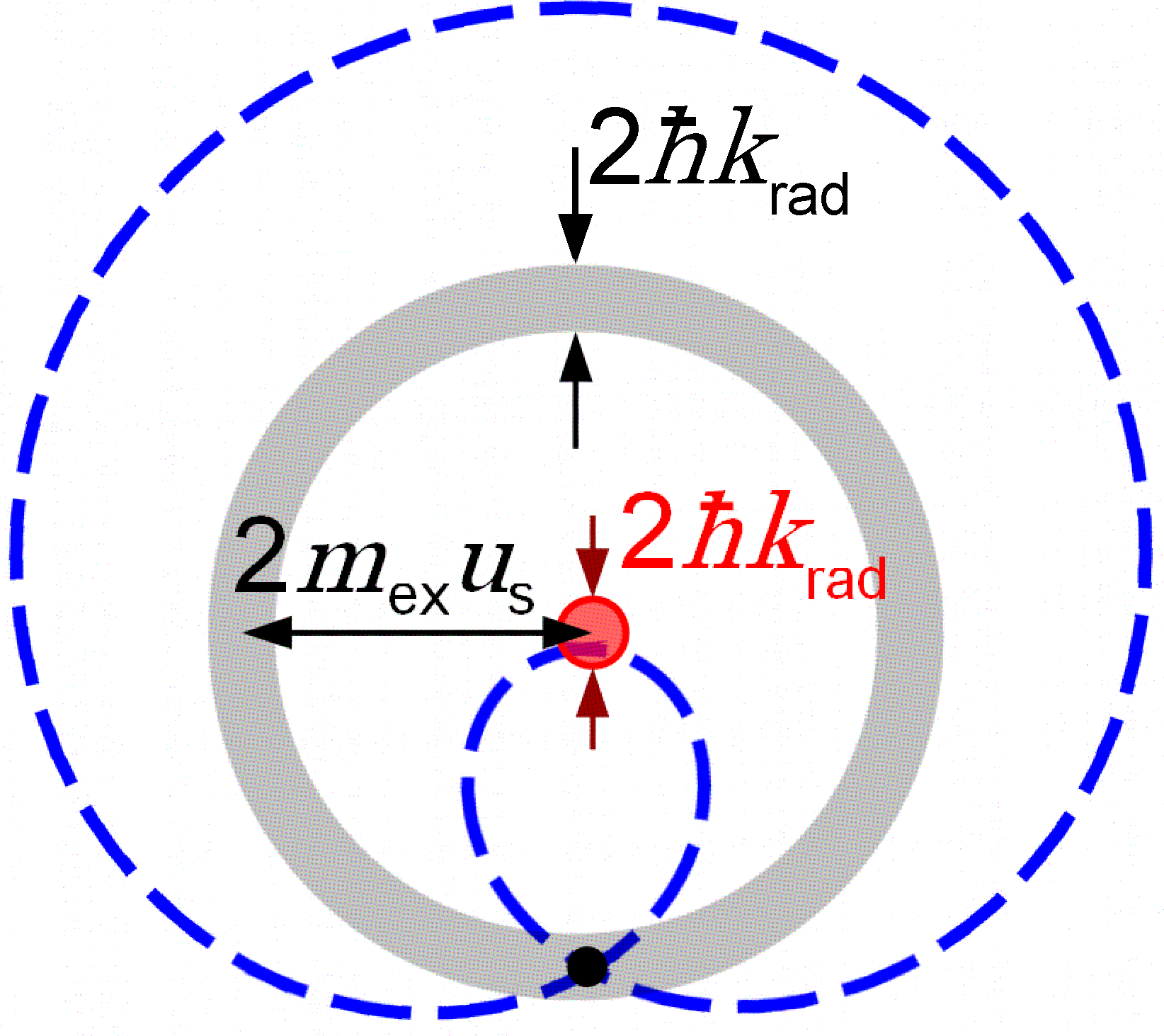}
\caption{\label{fig:conservation}(Color online)
Momenta $\vec{p}'$ satisfying the energy conservation
condition~(\ref{conservation=}).
For $\vec{p}$ in the narrow radiative region
$p<\hbar{k}_\mathrm{rad}$, shown by the small red circle,
the allowed $\vec{p}'$ lie in the narrow circular strip
$|p'-2m_\mathrm{ex}u_\mathrm{s}|<\hbar{k}_\mathrm{rad}$,
shown by the thick grey circular band.
For a given $\vec{p}$ in the strip, shown by a small black
circle, the allowed $\vec{p}'$ lie on a contour,
represented by the dashed blue line. Most of this contour
lies outside the radiative region.
}
\end{figure}

In turn, from which states can excitons be scattered
into the circular strip? If 
$|\vec{p}|=2m_\mathrm{ex}u_\mathrm{s}$, the set of all
momenta $\vec{p}'$ satisfying the energy conservation
condition~(\ref{conservation=}) forms a contour shown in
Fig.~\ref{fig:conservation} by the dashed line. Crucially,
most of the contour lies outside the radiative region. Thus,
the population of exciton states in the strip
$|p'-2m_\mathrm{ex}u_\mathrm{s}|<\hbar{k}_\mathrm{rad}$
is determined by exchanging excitons with states in a
broad energy interval, determined either by the largest
allowed phonon energy, $4m_\mathrm{ex}u_\mathrm{s}^2$,
or by the temperature~$T$, whichever is smaller. In either
case, this cutoff energy is much larger than that of the
radiative window,
$(\hbar{k}_\mathrm{rad})^2/(2m_\mathrm{ex})$.
Thus, even if one sets $f_\vec{p}=0$ in all the radiative
region (the extreme case of very fast radiative decay), the
effect on $f_\vec{p'}$ in the strip
$|p'-2m_\mathrm{ex}u_\mathrm{s}|<\hbar{k}_\mathrm{rad}$
is small by a factor
$\hbar{k}_\mathrm{rad}/
\min\{\sqrt{m_\mathrm{ex}T},m_\mathrm{ex}u_\mathrm{s}\}\ll{1}$,
Therefore, for $\vec{p}$ in the radiative region,
$I^\mathrm{in}_\vec{p}$ can be evaluated using
$f_\vec{p'}^\mathrm{eq}$.
This is the key observation that enables us to solve the
problem.

Then, looking for a solution $f_\vec{p}(t)=f_\vec{p}e^{-\gamma{t}}$
and setting $\partial{f}_\vec{p}/\partial{t}=-\gamma{f}_\vec{p}$
in Eq.~(\ref{kinetic=}), we readily find the exciton distribution
in the radiative region,
\begin{equation}\label{fp=}
f_\vec{p}=
\frac{I^\mathrm{in}_\vec{p}}{\gamma+\Gamma_\vec{p}+1/\tau_\vec{p}},
\end{equation}
where $I^\mathrm{in}_\vec{p}=f_\vec{p}^\mathrm{eq}/\tau_\vec{p}$
is fixed by the detailed balance condition. Furthermore,
in the radiative region we can neglect the momentum dependence
of $f_\vec{p}^\mathrm{eq}$ and $\tau_\vec{p}\approx\tau_0$.
Then, integrating over $\vec{p}$, we obtain
the following simple expression for the effective decay rate:
\begin{equation}\label{avGamma=}
\gamma=\frac{2\pi\hbar^2}{m_\mathrm{ex}T}
\int\frac{d^2\vec{p}}{(2\pi\hbar)^2}\,
\frac{\Gamma_\vec{p}}{1+\Gamma_\vec{p}\tau_0+\gamma\tau_0}.
\end{equation}
In fact, $\gamma\tau_0$ in the denominator can be safely
neglected, since from Eq.~(\ref{avGamma=}) it follows that
$\gamma\tau_0\leqslant
\hbar^2{k}_\mathrm{rad}^2/(2m_\mathrm{ex}T)$.
Since $\tau_\vec{p}\sim\tau_0$ in the thermal region of momenta,
the condition $\gamma\tau_\vec{p}\ll{1}$ is also satisfied, which
justifies that the distribution at $p>\hbar{k}_\mathrm{rad}$ is
thermal. Thus, these states indeed act as a quasistationary
reservoir, from which the population is supplied to the radiative
region.
If the exciton-phonon relaxation is not fast enough,
$1/\tau_0\ll\Gamma_\vec{p}$, then $\Gamma_\vec{p}$ drops
out of Eq.~(\ref{avGamma=}), and the overall decay rate is
governed by the exciton-phonon relaxation bottleneck.

In the following, we use the standard expressions for the
radiative decay rates of the longitudinal and transverse
excitons~\cite{Andreani1991,Divergence},
\begin{subequations}\begin{eqnarray}
&&\Gamma^\mathrm{L}_\vec{p}= \Gamma^\mathrm{vac}_0\,
\theta(\varepsilon\hbar^2\omega_\mathrm{ex}^2-c^2p^2)\,
\frac{\sqrt{\varepsilon\hbar^2\omega_\mathrm{ex}^2-c^2p^2}}%
{\vep\hbar\omega_\mathrm{ex}},\label{GammaL=}\\
&&\Gamma^\mathrm{T}_\vec{p}= \Gamma^\mathrm{vac}_0\,
\theta(\varepsilon\hbar^2\omega_\mathrm{ex}^2-c^2p^2)\,
\frac{\hbar\omega_\mathrm{ex}}%
{\sqrt{\varepsilon\hbar^2\omega_\mathrm{ex}^2-c^2p^2}},\label{GammaT=}
\end{eqnarray}\end{subequations}
where $\Gamma^\mathrm{vac}_0$ is the parameter characterizing
the exciton-photon coupling strength, determined by the
excitonic transition dipole moment. Then, the integral in
Eq.~(\ref{avGamma=}) can be straightforwardly evaluated:
\begin{subequations}\label{avGammaNarrow=}\begin{eqnarray}
\label{avGammaLT=}
&&\gamma^\mathrm{L,T}=
\frac{(\hbar\omega_\mathrm{ex})^2}{m_\mathrm{ex}c^2T}\,
\Gamma^\mathrm{vac}_0\sqrt\vep\,\mathcal{F}_\mathrm{L,T}\!
\left(\frac{\Gamma^\mathrm{vac}_0\tau_0(T)}{\sqrt\vep}\right),\\
\label{FL=}
&&\mathcal{F}_\mathrm{L}(x)\equiv\frac{1}{2x}-\frac{1}{x^2}
+\frac{\ln(1+x)}{x^3},\\
\label{FT=}
&&\mathcal{F}_\mathrm{T}(x)\equiv{1}-x\ln\left(1+\frac{1}x\right),
\end{eqnarray}\end{subequations}
where we included explicitly the temperature argument of
$\tau_0(T)$, to emphasize that it determines the temperature
dependence of $\gamma$, together with the
prefactor $1/T$.

The dependence $\tau_0(T)$ is determined by the specific
exciton-phonon coupling mechanism. We assume the main
mechanism to be the deformation potential arising from the
difference between conduction and valence band shifts under
a local deformation of the crystal. We neglect the
piezoelectric potential, which is due to the macroscopic
electric polarization created by the deformation. Indeed,
as the exciton is overall neutral, it can couple to an
electric field only in the second order (Stark effect).
For the deformation potential, we take a simple form
assuming the phonons to be two-dimensional, with wave
vectors~$q$ much smaller than the exciton radius:
\begin{equation}\label{Vdeform=}
\hat{V}(\vec{r})=(D_\mathrm{c}-D_\mathrm{v})\sum_\vec{q}
\sqrt{\frac{\hbar{q}}{2\rho{S}u_\mathrm{s}}}\,
\left(\hat{b}_{\vec{q}}+\hat{b}_{-\vec{q}}^{\dagger}\right)
e^{i\vec{q}\vec{r}},
\end{equation}
where $D_\mathrm{c,v}$ is the deformation potential for the
conduction/valence band,
$\rho$ and $S$ are the surface mass density and the total area
of the excitonic layer (so that $\rho{S}$ is the sample mass),
and $\hat{b}_{\vec{q}}^{\dagger},\,\hat{b}_{\vec{q}}$ are the
creation and annihilation operators for a longitudinal acoustic
phonon with wave vector~$\vec{q}$.
The Fermi Golden Rule gives the following phonon absorption
rate~\cite{Hanamura1988,FastPhonons}:
\begin{subequations}\begin{align}\label{tau0=}
&\frac{1}{\tau_0}=
\frac{2(D_\mathrm{c}-D_\mathrm{v})^2m_{\mathrm{ex}}^2}%
{\rho\hbar^3(e^{2m_{\mathrm{ex}}u_\mathrm{s}^2/T}-1)}\approx
\mathcal{A}\,\frac{T}\hbar\quad(T\gg m_{\mathrm{ex}}u_\mathrm{s}^2),\\
&\mathcal{A}\equiv\frac{(D_\mathrm{c}-D_\mathrm{v})^2m_{\mathrm{ex}}}%
{\rho\hbar^2{u}_\mathrm{s}^2}.\label{Upsilon=}
\end{align}\end{subequations}
Eq.~(\ref{tau0=}), in combination with
Eqs.~(\ref{avGammaLT=})--(\ref{FT=}), determines the temperature
dependence of the radiative relaxation rate for the whole exciton
population, both in the regime of full thermalization,
$1/\tau_0\gg\Gamma$, and for strong depletion of the radiative
zone due to relaxation bottleneck, $1/\tau_0\ll\Gamma$, which
inevitably sets in at low temperatures.

In semiconductor quantum wells, the simple model used so far
needs several modifications.
The most important one is that while the exciton motion is
confined to two dimensions, the phonons are three-dimensional,
so their wave vector has also a perpendicular component~$q_z$
in addition to the in-plane~$\vec{q}$.
Also, one often has to take into account the finite values of
the exciton radius $a_\mathrm{ex}$ and of the quantum well
thickness~$d_\mathrm{QW}$.
The electron-phonon matrix element then also includes factors
which vanish for large $q\gg 1/a_\mathrm{ex}$ and
$|q_z|\gg 1/d_\mathrm{QW}$~\cite{Takahagara1985}.
For most III-V and II-VI semiconductors, such as GaAs or ZnSe,
the heavy-hole exciton mass $m_\mathrm{ex}\sim{m}_0$, the free
electron mass, and $u_\mathrm{s}\sim{5}\times{10}^5\:\mbox{cm/s}$,
which gives the phonon wave vector satisfying the energy
conservation~(\ref{conservation=}),
$2m_\mathrm{ex}u_\mathrm{s}/\hbar\sim0.01\:\mbox{\AA}^{-1}$.
The typical values of $1/a_\mathrm{ex}$ and $1/d_\mathrm{QW}$
are often of the same order.

Then, phonons with more or less any $|q_z|$ on the scale of
Fig.~\ref{fig:conservation} can be emitted or absorbed, so
the radiative region can be supplied not just from the narrow
strip around $p'=2m_\mathrm{ex}u_\mathrm{s}$, but from a wide
region in the outer space $p'>2m_\mathrm{ex}u_\mathrm{s}$.
A state with $\vec{p}$ in the outer region, in turn, can receive
population from the whole inner area of the inner part of the
dashed contour (phonon absorption), and from the whole outer
area of its outer part (phonon emission). Thus,
population depletion in the small radiative region still only
weakly affects the supply region, so our key observation that
$I^\mathrm{in}_\vec{p}$ can be calculated using
$f_\vec{p}^\mathrm{eq}$ for the termal states remains valid,
together with
Eqs.~(\ref{avGamma=}) and (\ref{avGammaLT=})--(\ref{FT=}).
What should be modified with respect to the simple two-dimensional
model, is the phonon absorption rate $1/\tau_0(T)$.
While the Fermi Golden Rule calculation with the simple
Hamiltonian~(\ref{Vdeform=}) gives a rate $1/\tau_0\propto{T}^2$,
the model including the matrix element suppression at
$q\gg 1/a_\mathrm{ex}$ and $|q_z|\gg 1/d_\mathrm{QW}$, yields
$1/\tau_0\propto{T}$~\cite{Piermarocchi1996}.
Both lead to relaxation bottleneck at low temperatures;
more detailed investigation of this issue is beyond the
scope of this paper.

\section{Application to transition metal dichalcogenides}

Exciton relaxation by scattering on acoustic phonons
in monolayer TMDCs, such as MoS$_2$, MoSe$_2$, WS$_2$,
WSe$_2$ was recently studied in Ref.~\cite{Thilagam2016},
where three-dimensional phonons were considered.
Here we assume the binding between the TMDC monolayer and
the substrate not to be strong, so the acoustic phonons
are taken to be two-dimensional.
To extend the validity of Eq.~(\ref{Vdeform=}) to phonon
wave vectors comparable to inverse exciton radius,
$1/a_\mathrm{ex}$, the
matrix element should be multiplied by the Fourier transform
of the square of the excitonic wave function for the
electron-hole relative coordinate~$\vec{r}_\mathrm{eh}$.
For the hydrogen-like wave function,
$\propto{e}^{-r_\mathrm{eh}/a_\mathrm{ex}}$, and equal
electron and hole masses $m_\mathrm{ex}/2$, this amounts
to an additional factor $[1+(qa_\mathrm{ex}/4)^2]^{-3/2}$
in Eq.~(\ref{Vdeform=}).
It is known that because of strong dielectric confinement
in the TMDC monolayer, the interaction potential is not
$1/r_\mathrm{eh}$, so the bound state wave functions do not
have a hydrogenic form~%
\cite{Berkelbach2013,Qiu2013,Chernikov2014,Ye2014,He2014}.
However, as will be seen below, the precise form of the
cutoff factor does not matter, as the typical exciton radius
in TMDC, $a_\mathrm{ex}\sim{1}\:\mbox{nm}$~%
\cite{Ugeda2014,Stier2016}, will turn out
to be small enough for the cutoff effect not to play a
significant role.

Another modification of the simple model studied in the
previous section concerns the variety of excitonic
species in TMDCs.
The longitudinal and transverse
bright excitons, degenerate at $\vec{p}=0$, become
strongly split by the exchange interaction at
$p\gg\hbar{k}_\mathrm{rad}$~%
\cite{Glazov2014,Gartstein2015,Yu2014}.
Namely, in addition to $p^2/(2m_\mathrm{ex})$, the longitudinal
exciton energy contains a linear term, $v_\mathrm{ex}p$,
with the group velocity,
$v_\mathrm{ex}=c\Gamma_0^\mathrm{vac}/(2\vep\omega_\mathrm{ex})$,
determined by the same parameter $\Gamma_0^\mathrm{vac}$
as the radiative rates~(\ref{GammaL=}), (\ref{GammaT=}).
Using the parameters from Table~\ref{tab:values} and taking
$\vep=2.5$, we obtain
$v_\mathrm{ex}=(0.9-1.1)\times{10}^7\:\mbox{cm/s}$ for
all four materials. Not only
$v_\mathrm{ex}\gg{u}_\mathrm{s}$,
but also the energy $m_\mathrm{ex}v_\mathrm{ex}^2$,
at which $v_\mathrm{ex}p$ is overcome by
$p^2/(2m_\mathrm{ex})$, is
2--3 times larger than the room temperature.
Thus, at low tempratures, the longitudinal exciton
population is negligible compared to the transverse one.
The matrix element of the deformation
potential~(\ref{Vdeform=}), which
determines the rate $W_{\vec{p}'\to\vec{p}}$, should
also include the overlap
$\langle\alpha,\vec{p}|\alpha',\vec{p}'\rangle$
between states of the two exciton branches,
$\alpha,\alpha'=\mathrm{L},\mathrm{T}$,
determined by the angle $\phi_{\vec{p}\vec{p}'}$
between $\vec{p}$ and $\vec{p}'$~\cite{Gartstein2015,Yu2014}):
\begin{equation}\label{overlaps=}
\left.\begin{array}{c}
|\langle\mathrm{T},\vec{p}|\mathrm{T},\vec{p}'\rangle|^2\\
|\langle\mathrm{L},\vec{p}|\mathrm{T},\vec{p}'\rangle|^2
\end{array}\right\}=\frac{1\pm\cos(2\phi_{\vec{p}\vec{p}'})}{2}.
\end{equation}

\begin{table}
\begin{center}\begin{tabular}{|c|c|c|c|c|c|}
\hline
{}  & MoS$_2$ & MoSe$_2$ & WS$_2$ & WSe$_2$ & units \\ \hline
$\hbar\omega_\mathrm{ex}$  & 1.9  & 1.7 & 2.0 & 1.7 & eV \\
$1/\Gamma^\mathrm{vac}_0$  & 0.23  & 0.24 & 0.19 & 0.22 & ps \\
$m_\mathrm{ex}$  & $1.09$  & $1.35$ & $0.73$ & $0.90$ & $m_0$ \\
$D_\mathrm{c}-D_\mathrm{v}$  & 2.0   & 0.6  & 1.5  & 1.1  & eV \\
$u_\mathrm{s}$ & $6.6$   & $4.1$  & $4.3$  & $3.3$  & $10^5\:\mathrm{cm/s}$ \\
$\rho$  & 1.6 & 2.0 & 2.4 & 3.1  & $10^{-7}\:\mbox{g/cm}^2$ \\
\hline
\end{tabular}\end{center}
\caption{\label{tab:values}
Values of the TMDC parameters used in the estimates.
$\hbar\omega_\mathrm{ex}$ and $1/\Gamma^\mathrm{vac}_0$ are taken from
Ref.~\cite{Palummo2015},
$m_\mathrm{ex}$, $D_\mathrm{c}-D_\mathrm{v}$, $u_\mathrm{s}$ from
Ref.~\cite{Jin2014}, $\rho$~from Ref.~\cite{Thilagam2016},
and $m_0$~is the free electron mass. }
\end{table}

Besides the two bright exciton species, in monolayer TMDCs there are
six dark exciton species whose energies are in the same range (up to
energy shifts of a few tens of meV, due to spin-orbit and exchange
interactions).
These excitons are formed by conduction and valence band states
belonging to different valleys or/and having different spins,
so their radiative decay is forbidden in the zero approximation.
In molybdenum-based compounds, the bright excitons have a lower
energy than the dark ones, while in tungsten-based ones, the
situation is the opposite. This was used to explain the experimentally
observed rise of the luminescence intensity in WSe$_2$ with increasing
temperature in terms of the increasing thermal population of the
higher-energy bright excitons~%
\cite{Arora2015,Withers2015,Wang2015,Zhang2015}.
Conversion between dark and bright excitons requires either a spin
flip or an intervalley scattering which, in a clean crystal, can
occur by phonon absorption or emission.
The energy of intervalley phonons is of the same order as that of
optical phonons, so these processes are slow at low temperature,
as compared to scattering by acoustic phonons.
Thus, conversion between dark and bright excitons leads to slow
equilibration between different excitonic reservoirs but does not
affect directly the population of the radiative states.
In the following, we focus on the dynamics of exciton exchange
between the bright radiative states and the reservoir from the
transverse bright exciton band, leaving aside the problem of
slow population exchange between different reservoirs. Still, one
should keep in mind that this slow population exchange may
contribute to dynamics at very long times.


For bright excitons in TMDCs, the radiative rate is quite large,
$\hbar\Gamma_0^\mathrm{vac}$ being a few meV~
\cite{Moody2015, Poellmann2015, Jakubczyk2016,%
Palummo2015,Selig2016, Wang2016}.
For the coefficient $\mathcal{A}$~in
Eqs.~(\ref{tau0=}), (\ref{Upsilon=}), the material parameters
listed in Table~\ref{tab:values} give a few $\mu\mbox{eV/K}$.
This is an order of magnitude smaller than experimentally
measured values~%
\cite{Moody2015,Koirala2016,Dey2016,Selig2016,Jakubczyk2016},
probably due to a larger value of $D_\mathrm{c}-D_\mathrm{v}$
than that given in Ref.~\cite{Jin2014}.
In any case, $1/\tau_0(T)<\Gamma_0^\mathrm{vac}$ below several
tens of Kelvins, so one can expect the bottleneck effect.
It should be noted, that intravalley dark excitons can also
decay radiatively at a rate about 100--1000 times slower than
for the bright ones~\cite{Slobodeniuk2016}, so already above
a few Kelvins, their population in the radiative region
should be thermalized. For intervalley dark excitons, radiative
decay is possible if assisted by phonon emission~\cite{Dery2015}.
We are not aware of any estimate for the rate of this process.

\begin{figure}
\includegraphics[width=8cm]{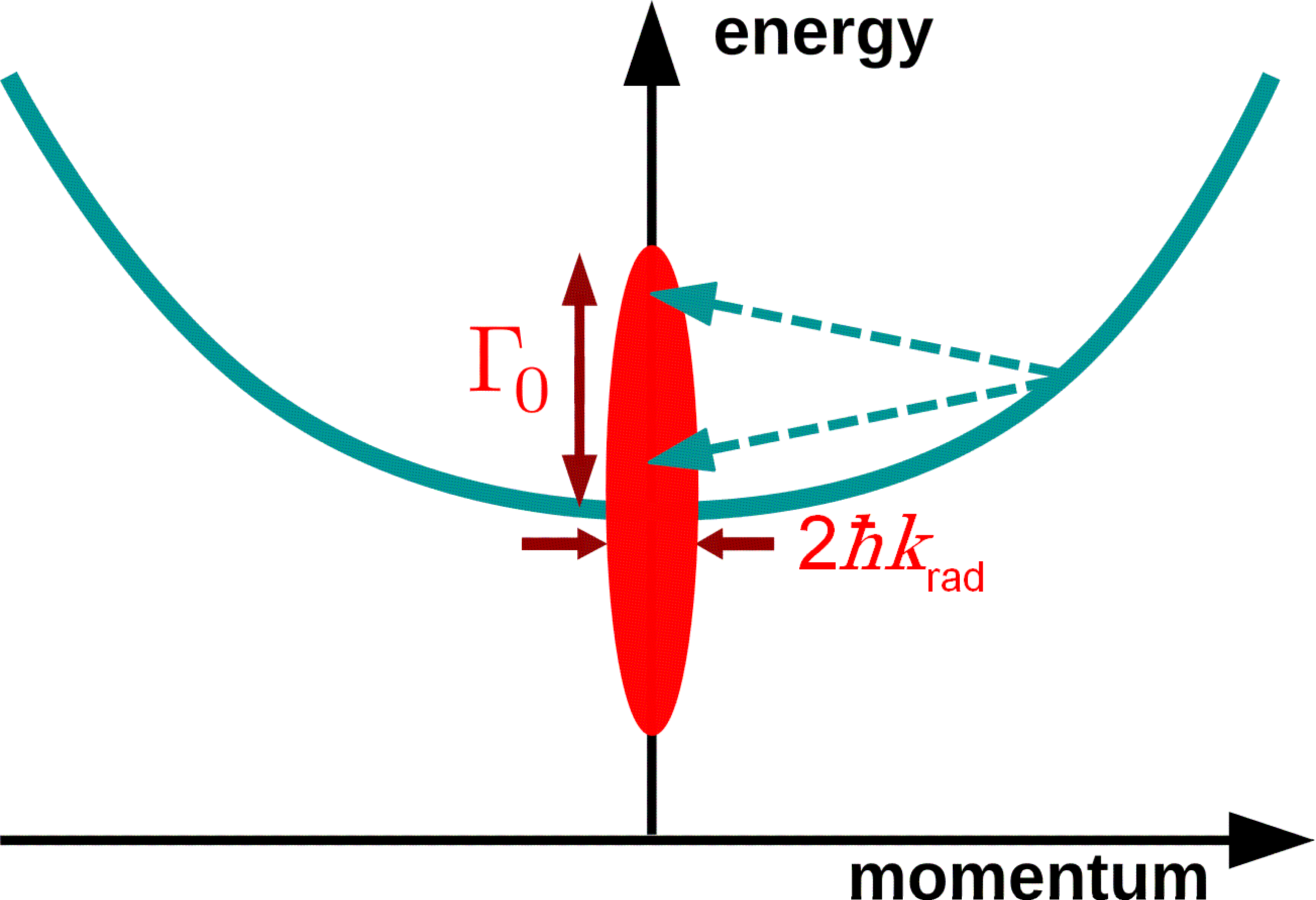}
\caption{\label{fig:broadening}(Color online)
A schematic representation of the exciton band (solid curve).
The radiative states, strongly broadened in energy (but not in
momentum) are shown by the red ellipse.
The dashed arrows show exciton transitions from the reservoir
to the radiative states, accompanied by phonon absorption and
emission (the slope of the arrows corresponds to the speed of
sound).
}
\end{figure}

The large radiative decay rate~$\Gamma_\vec{p}$ of the bright
excitons leads to an energy broadening
$\hbar\Gamma_0^\mathrm{vac}
\gg 2m_\mathrm{ex}u_\mathrm{s}^2$, the latter being 0.54,
0.27, 0.15, and 0.11~meV, for the four materials,
respectively. This introduces a large energy uncertainty
$\sim\hbar\Gamma_\vec{p}$ in Eq.~(\ref{conservation=}),
so the image of Fig.~\ref{fig:conservation} is not valid,
and the radiative region is refilled from states $\vec{p}'$
with energies
$(p')^2/(2m_\mathrm{ex})\sim\min\{\hbar\Gamma,T\}$
(Fig.~\ref{fig:broadening}). Still,
outside the radiative region the states are not broadened,
so the kinetic equation is applicable, and the effect of
the depleted radiative region on $f_{\vec{p}'}$ at energies
$(p')^2/(2m_\mathrm{ex})\sim\min\{\hbar\Gamma,T\}$ is
relatively weak by the same phase-space argument as before
(note that the total spectral weight of the radiative states
is unchanged, it is only spread over a wide energy range).
Thus, the population flow $I^\mathrm{in}_\vec{p}$ into the
radiative region can again be calculated assuming
equilibrium distribution $f_{\vec{p}'}^\mathrm{eq}$
outside, but not the detailed balance condition.
Indeed, scattering of an exciton from the non-radiative
region into a strongly broadened radiative state
followed by fast radiative decay can equivalently be
viewed as phonon-assisted radiative decay of a non-radiative
state via a virtual intermediate state in the
radiative region. Assuming
$m_\mathrm{ex}u_\mathrm{s}^2,\hbar/\tau_0\ll(\hbar\Gamma,T)\ll
m_\mathrm{ex}v_\mathrm{ex}^2$~\cite{FastSlowPhonons},
we can determine the average decay rate from the
total incoming flux into the radiative region for the
longitudinal and transverse excitons:
\begin{eqnarray}\label{avGammaIin=}
&&\gamma=\int\limits_{p<\hbar{k}_\mathrm{rad}}
\frac{d^2\vec{p}}{(2\pi\hbar)^2}\,
\frac{I^\mathrm{in,L}_\vec{p}+I^\mathrm{in,T}_\vec{p}}{n_\mathrm{ex}}.
\end{eqnarray}
The incoming fluxes
$I^\mathrm{in,L}_\vec{p},I^\mathrm{in,T}_\vec{p}$ are, in turn,
given by (we omit the labels ``L,T'' at $I^\mathrm{in}_\vec{p}$
and $\Gamma_\vec{p}$ for brevity)
\begin{eqnarray}
&&I^\mathrm{in}_\vec{p}=\frac{n_\mathrm{ex}}{2}
\int\limits_{\hbar{k}_\mathrm{rad}}^\infty
\frac{p'\,dp'}{m_\mathrm{ex}{T}}\,
\frac{e^{-\ep_{p'}/T}(D_\mathrm{c}-D_\mathrm{v})^2p'}{2\rho{u}_\mathrm{s}
[1+(p'/\hbar)^2(a_\mathrm{ex}/4)^2]^3}
\times\nonumber\\
&&\qquad{}\times\sum_\pm
\frac{e^{\mp{u}_\mathrm{s}p'/(2T)}}%
{2\sinh[u_\mathrm{s}p'/(2T)]}\,
\frac{\Gamma_\vec{p}}%
{(\ep_{p'}\pm{u}_\mathrm{s}p')^2+\hbar^2\Gamma_\vec{p}^2/4},
\nonumber\\
\label{IinLorentzian=}
\end{eqnarray}
where we approximated $|\vec{p}-\vec{p}'|\approx{p}'$,
took into account the overlaps~(\ref{overlaps=})
which amount to $1/2$ upon angular integration, and
denoted $\ep_{p'}\equiv(p')^2/(2m_\mathrm{ex})$. The sum over
the two signs in the second line corresponds to phonon
absorption/emission for the upper/lower signs, respectively.
The key ingredient of
Eq.~(\ref{IinLorentzian=}) is the Lorentzian density of final
states for exciton scattering into the radiative region,
which includes the radiative broadening.
The $p'$~integral is cut off at large $p'$ by one of the
three factors: the thermal exponential $e^{-\ep_{p'}/T}$,
the matrix element suppression
$[1+(p'/\hbar)^2(a_\mathrm{ex}/4)^2]^{-3}$, and the
Lorentzian which imposes $\ep_{p'}\lesssim\Gamma_\vec{p}$.
First, all of these imply $u_\mathrm{s}p'\ll\ep_{p'},T$,
as illustrated by Fig.~\ref{fig:broadening}, so we
approximate
$2\sinh[u_\mathrm{s}p'/(2T)]\approx u_\mathrm{s}p'/T$ and
neglect $u_\mathrm{s}p'$ everywhere else, which is equivalent
to treating the phonon-induced potential $V(\vec{r})$ as a
quasistatic disorder of the strength
$\langle{V}^2\rangle\propto{T}$.
Second, the cutoff imposed by the Lorentzian turns out to be
more important than that due to the exciton radius: for
$\ep_{p'}=3\:\mbox{meV}\sim\hbar\Gamma_\vec{p}$, $m_\mathrm{ex}=m_0$,
$a_\mathrm{ex}=1\:\mbox{nm}$, we obtain
$p'a_\mathrm{ex}/(4\hbar)\approx 0.07$.
Thus, the integral is dominated by $\ep_{p'}\sim\hbar\Gamma$ for
$T\gg\hbar\Gamma$, and by $\ep_{p'}\sim{T}$ for $T\ll\hbar\Gamma$,
while the cutoff due to the exciton radius is not important in
either case.

\begin{figure}
\includegraphics[width=7cm]{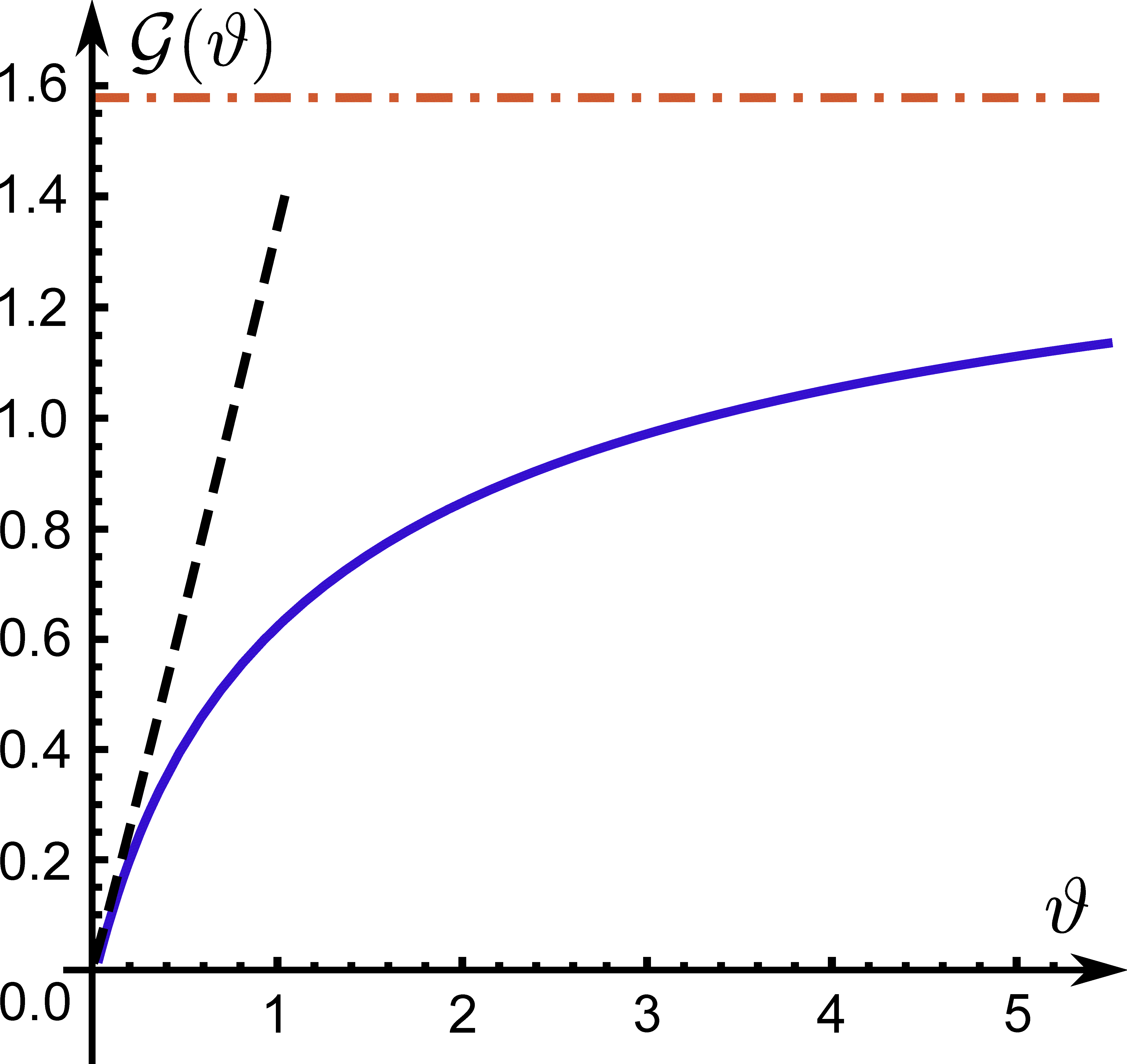}
\caption{\label{fig:Gfunction}(Color online)
The function $\mathcal{G}(\vartheta)$, defined in Eq.~(\ref{Gtheta=}),
which represents the dimensionless temperature dependence of
$\gamma$. The asymptotic behaviour is
$\mathcal{G}(\vartheta\to{0})\sim(4/3)\vartheta$ (dashed black line),
and $\mathcal{G}(\vartheta\to\infty)=\pi/2$ (dot-dashed red line).
}
\end{figure}

Combining Eqs.~(\ref{avGammaIin=}), (\ref{IinLorentzian=}),
and integrating first over $\vec{p}$, then over $\vec{p}'$,
we obtain
\begin{subequations}\begin{align}\label{avGammaBroad=}
&\gamma=\mathcal{A}\,\frac{\hbar{k}_\mathrm{rad}^2}{2\pi{m}_\mathrm{ex}}\,
\mathcal{G}\!\left(\frac{2 T\sqrt\vep}{\hbar\Gamma_0^\mathrm{vac}}\right),\\
&\mathcal{G}(\vartheta)\equiv\int\limits_0^\infty
\left(1-x\arctan\frac1x+\frac{x-\arctan{x}}{x^3}\right)e^{-x/\vartheta}\,dx,\label{Gtheta=}
\end{align}\end{subequations}
with the same coefficient $\mathcal{A}$  as in Eqs.~(\ref{tau0=}), (\ref{Upsilon=}).
The function $\mathcal{G}(\vartheta)$, defined in Eq.~(\ref{Gtheta=})
and plotted in Fig.~\ref{fig:Gfunction}, determines the temperature
dependence of~$\gamma$ in the appropriate units:
for $\vep=2.5$, we obtain
$\hbar\Gamma_0^\mathrm{vac}/(2\sqrt\vep)\sim{10}\:\mbox{K}$,
while
$\hbar{k}_\mathrm{rad}^2/(2\pi{m}_\mathrm{ex})\approx{5}\:\mbox{ns}^{-1}$
(taking $m_\mathrm{ex}=m_0$, $\hbar\omega_\mathrm{ex}=2\:\mbox{eV}$).
The dimensionless coefficient $\mathcal{A}$ is in the range $0.03-0.13$
for the parameters in Table~\ref{tab:values}, while in experiments
values approaching unity are reported~%
\cite{Moody2015,Koirala2016,Dey2016,Selig2016,Jakubczyk2016}.
Thus, the typical values of $\gamma$ at temperatures $T\sim 50\:\mathrm{K}$
correspond to decay times of the order of nanoseconds.

In time-resolved photoluminescence experiments on TMDCs,
two contributions to the luminescence
are often observed~\cite{Korn2011,Lagarde2014,Robert2016}:
a fast component, decaying on the picosecond time scale, and a slow
one, which appears at temperatures above 100--150~K, and
decays on time scales between 100~ps~\cite{Korn2011}
and 2.5~ns~\cite{Robert2016}.
It is this slow component that was attributed to the radiative
decay of excitons from the reservoir, and the nanosecond decay time
scale is similar to what we obtained above.
At low temperatures, the slow component was not observed so far.
It is likely that in the above experiments at low temperatures,
the excitons quickly decay from the radiative states without
populating the non-radiative reservoir at all. This is quite
natural for quasiresonant optical excitation used in
Refs.~\cite{Lagarde2014,Robert2016}, but not totally clear for
the non-resonant excitation in Ref.~\cite{Korn2011}.

The overall temperature dependence of $\gamma$, given by
Eqs.~(\ref{avGammaBroad=}), (\ref{Gtheta=}), has quite simple
qualitative explanation.
At $T\ll\hbar\Gamma_0^\mathrm{vac}$,
we have $\gamma\propto{T}/\Gamma_0^\mathrm{vac}$.
Indeed, the whole thermal exciton population pumps the
radiative region, where the density of states is
$\propto{1}/\Gamma_0^\mathrm{vac}$ (the top of the Lorentzian),
while the phonon occupation is proportional to~$T$.
At $T\gg\hbar\Gamma_0^\mathrm{vac}$, the whole Lorentzian in the
radiative region is involved, which can be pumped only from
states with energies $\ep_{p'}\sim\hbar\Gamma_0^\mathrm{vac}$
whose population is $\propto{1}/T$, while the pumping rate is
proportional to $1/\tau_0(T)=\mathcal{A}{T}/\hbar$, so the
temperature drops out. Note that Eq.~(\ref{IinLorentzian=})
is valid only when $\Gamma_\vec{p}\gg{1}/\tau_\vec{p}$~%
\cite{FastSlowPhonons}, which sets the upper limit
on the temperature to be a few tens of Kelvins.
In particular, if $\mathcal{A}\sim{1}$, there is no room for
the temperature-independent regime.
At higher temperatures, when
$\Gamma_\vec{p}\lesssim{1}/\tau_\vec{p}$,
the rate $\gamma$ should start decreasing with temperature,
but the study of this regime is beyond the scope of the present
paper.

\section{Conclusions}

To conclude, we analyzed radiative decay of thermalized excitonic
population subject to scattering by acoustic phonons. Assuming
sufficiently low temperatures, we did not include scattering on
optical phonons.
Depending on the radiative decay rate~$\Gamma_0$ of excitonic
states with small momenta, on the phonon absorption rate
$1/\tau_0$ in these states, and on the frequency
$2m_\mathrm{ex}u_\mathrm{s}^2/\hbar$ of the absorbed phonons,
we identified several regimes.
When $\Gamma_0\ll{1}/\tau_0,m_\mathrm{ex}u_\mathrm{s}^2/\hbar$,
the excitons in the radiative region are thermalized, and the
overall population decay rate $\gamma\propto{1}/T$, determined
by the thermal population of the radiative
region~\cite{Andreani1991}.
When $1/\tau_0\ll\Gamma_0$, the radiative region is strongly
depleted, and $\gamma$~is determined by exciton scattering into
the radiative region (the relaxation bottleneck).
In this case, $\Gamma_0$ drops out from~$\gamma$, and its
temperature dependence is $\gamma\propto{1}/(T\tau_0(T))$.
In particular, if $1/\tau_0\propto{T}$, a curious situation
may arise where $\gamma$ depends neither on temperature, nor
on $\Gamma_0$.
Finally, when
$\Gamma_0\gg{1}/\tau_0,m_\mathrm{ex}u_\mathrm{s}^2/\hbar$,
the radiative broadening of states in the depleted radiative
region affects the exciton-phonon scattering rate itself.
It is this strong-broadening case that we find to be relevant
for bright excitons in TMDCs at low temperatures.
Then, at lowest temperatures $T\ll\hbar\Gamma_0$ the effective
population decay rate $\gamma\propto{T}$, while at
$T\gg\hbar\Gamma_0$ it becomes temperature-independent.

\section{Acknowledgements}
We thank M.~Potemski and T.~Jakubczyk for stimulating discussions.
A. O. S. acknowledges financial support from
the EC Graphene Flagship project (No. 604391).


\begin{thebibliography}{99}
\bibitem{FastPhonons}
At sufficiently high temperatures, $\hbar/\tau_0(T)$ becomes
of the order of $2m_\mathrm{ex}u_\mathrm{ex}^2$ or higher.
Then, the Fermi Golden Rule used to derive Eq.~(\ref{tau0=})
is not applicable any more.
One way to go beyond the Golden Rule is to use the
self-consistent Born approximation, which gives the same result
but with a different numerical factor. However, at energies
$p^2/(2m_\mathrm{ex})\sim\hbar/\tau_0$ the self-consistent
Born approximation is not valid quantitatively, as well as
the kinetic equation itself, so Eq.~(\ref{avGamma=}) is not
expected to be quantitatively valid unless
$\hbar/\tau_0\ll{m}_\mathrm{ex}u_\mathrm{ex}^2$.
\bibitem{FastSlowPhonons}
At this point, it does not matter whether
$m_\mathrm{ex}u_\mathrm{s}^2\gg\hbar/\tau_0$ or not. Indeed,
the main contribution to $I_\vec{p}^\mathrm{in}$ comes from
the states with relatively high energies,
$(p')^2/(2m_\mathrm{ex})\sim\{\hbar\Gamma,T\}\gg\hbar/\tau_0$,
so the kinetic equation at these high energies is valid.
However, at too high temperatures, when $1/\tau_0$ becomes
comparable to~$\Gamma$, our theory is no longer quantitatively
valid. Qualitatively, the effect can be taken into account
by replacing $\Gamma_\vec{p}\to\Gamma_\vec{p}+1/\tau_\vec{p}$
in the denominator of the Lorentzian in Eq.~(\ref{IinLorentzian=}).
However, the precise shape of the spectral function at the
bottom of the exciton band will no longer be Lorentzian.
\end{thebibliography}
\end{document}